\documentclass[%
 aip,
 amsmath,amssymb,
 reprint,%
]{revtex4-1}

\usepackage{graphicx}
\usepackage{dcolumn}
\usepackage{bm}
\usepackage{hyperref}

\usepackage[utf8]{inputenc}
\usepackage[T1]{fontenc}
\usepackage{mathptmx}

\usepackage{siunitx} 
\usepackage[version=4]{mhchem}
\usepackage{physics}

\hypersetup{
    colorlinks=true,
    citecolor=magenta,
    linkcolor=magenta,
}

\begin{document}

\preprint{AIP/123-QED}

\title[Cd acceptors in \ce{Ga2O3}, an atomistic view]{Cd acceptors in \ce{Ga2O3}, an atomistic view}

\author{M. B. Barbosa}
\email{marcelo.barbosa@fc.up.pt}
\altaffiliation[Also at]{
C2TN, DECN, Instituto Superior T\'{e}cnico, Universidade de Lisboa, Bobadela, Portugal
}%
\affiliation{
 IFIMUP and IN \--- Institute of Nanoscience and Nanotechnology, \\
 Departamento de F\'{i}sica e Astronomia da Faculdade de Ci\^{e}ncias da Universidade do Porto, \\ 
 Rua do Campo Alegre 687, 4169-007 Porto, Portugal
}%

\author{J. G. Correia}
\altaffiliation[Also at]{
 EP Department, European Organization for Nuclear Research (CERN), CH-1211 Geneva, Switzerland
}%
\affiliation{
C2TN, DECN, Instituto Superior T\'{e}cnico, Universidade de Lisboa, Bobadela, Portugal
}%

\author{K. Lorenz}
\affiliation{
 INESC-MN, IPFN, Instituto Superior T\'{e}cnico, Universidade de Lisboa, Lisboa, Portugal
}%

\author{A. S. Fenta}
\altaffiliation[Also at]{
 KU Leuven, Instituut voor Kern- en Stralingsfysica, Celestijnenlaan 200 D, 3001 Leuven, Belgium
}%
\altaffiliation{
EP Department, European Organization for Nuclear Research (CERN), CH-1211 Geneva, Switzerland
}%
\affiliation{
 Physics Department and CICECO, University of Aveiro, 3810-193 Aveiro, Portugal
 }%

\author{J. Schell}
\altaffiliation[Also at]{
EP Department, European Organization for Nuclear Research (CERN), CH-1211 Geneva, Switzerland
}%
\affiliation{
 Institute for Materials Science and Center for Nanointegration, Duisburg-Essen (CENIDE), University of Duisburg-Essen, 45141 Essen, Germany
}%

\author{R. Teixeira}
\affiliation{
C2TN, DECN, Instituto Superior T\'{e}cnico, Universidade de Lisboa, Bobadela, Portugal
}%

\author{E. Nogales}
\affiliation{
 Departamento de F\'{i}sica de Materiales, Universidad Complutense de Madrid, 28040 Madrid, Spain
}%

\author{B. M\'{e}ndez}
\affiliation{
 Departamento de F\'{i}sica de Materiales, Universidad Complutense de Madrid, 28040 Madrid, Spain
}%

\author{A. Stroppa}
\affiliation{
 CNR-SPIN c/o Universit\`{a} degli Studi dell'Aquila, Via Vetoio 10, I-67010 Coppito (L'Aquila), Italy
}%

\author{J. P. Ara\'{u}jo}
\affiliation{
 IFIMUP and IN \--- Institute of Nanoscience and Nanotechnology, \\
 Departamento de F\'{i}sica e Astronomia da Faculdade de Ci\^{e}ncias da Universidade do Porto, \\ 
 Rua do Campo Alegre 687, 4169-007 Porto, Portugal
}%

\date{\today}

\begin{abstract}
Finding suitable p-type dopants, as well as reliable doping and characterization methods for the emerging wide bandgap semiconductor $\beta$-\ce{Ga2O3} could strongly influence and contribute to the
development of the next generation of power electronics. In this work, we combine easily accessible ion implantation, diffusion and nuclear transmutation methods to properly incorporate the Cd dopant into the $\beta$-\ce{Ga2O3} lattice, being subsequently characterized at the atomic scale with the Perturbed Angular Correlation (PAC) technique and Density Functional Theory (DFT) simulations. The acceptor character of Cd in $\beta$-\ce{Ga2O3} is demonstrated, with Cd sitting in the octahedral Ga site in the negative charge state, showing no evidence of polaron deformations nor extra point defects nearby. Furthermore, thermally activated free electrons were observed for temperatures above $\sim$\SI{648}{\kelvin} with an activation energy of \SI{0.54+-0.01}{\electronvolt}. At lower temperatures the local electron transport is dominated by a tunneling process between defect levels and the Cd probe.
\end{abstract}



\maketitle

\ce{Ga2O3} is a wide band gap semiconductor of growing interest due to its potential application in power and high-voltage electronic devices \cite{Stepanov2016,Ratnaparkhe2017,Pearton2018}. It is transparent in the ultraviolet (UV) range, thus being also very promising for solar blind UV optoelectronic devices \cite{Varley2010,Lorenz2014,Varley2016,Pearton2018}.
With a band gap of 4.8 eV \cite{Tippins1965}, $\beta$-\ce{Ga2O3}, the most stable of its polymorphic forms, is a large gap insulator, but its conductivity actually depends on doping and growth conditions \cite{Stepanov2016,Pearton2018}. N-type semiconductivity is often observed and commonly attributed to ionized oxygen vacancies which act as donors \cite{Stepanov2016,Binet1998}, but Varley \emph{et al.} \cite{Varley2010,Varley2016} question this assumption attributing it to background impurities such as hydrogen, silicon and germanium, after first-principles calculations have revealed that oxygen vacancies act as deep donors in \ce{Ga2O3} and thus cannot be responsible for electron conductivity. Concerning p-type doping, several potential candidates have been proposed, such as Ti \cite{Tomm2001}, Cd \cite{Barbosa2013} and Mg \cite{Galazka2014}. In some cases, p-type conductivity has even been reported, but carrier mobility and concentrations have not been analyzed or revealed too low for practical applications \cite{Tomm2001,Liu2010}. On the other hand, theorists pointed out fundamental issues for p-type doping as potentially due to strong hole localization in \ce{Ga2O3} limiting hole mobility \cite{Varley2012}, and recent DFT studies indicate that such behavior is typical for cation-site impurities such as group 2 (e.g. Mg, Ca) and group 12 (e.g. Zn, Cd) acceptors which furthermore exhibit deep acceptor levels \cite{Lyons2018}. Moreover, the most studied acceptor, Mg, was shown to be incorporated not only in substitutional Ga-sites but also in interstitial sites as well as the substitutional O-site where it acts as compensating donor \cite{Peelaers2019}. On the other hand, a cathodoluminescence study on Zn-doped \ce{Ga2O3} suggests a relatively shallow acceptor level 0.26 eV above the valence band maximum \cite{Wang2014}. In the case of Cd, previous works report evidence of Cd located in substitutional Ga sites in $\beta$-\ce{Ga2O3} \cite{Pasquevich1993,Barbosa2013,Steffens2016}. Therefore, to better understand the concrete potentialities of p-type doping using Cd, this work is focused on the atomic scale study of location and charge state of implanted/diffused Cd probes in $\beta$-\ce{Ga2O3}, complemented by the study of electron mobility in this material. To address this problem, the Perturbed Angular Correlation technique (PAC) was used \cite{Wichert1999}. It accurately measures the hyperfine interaction between the nuclear quadrupole moment (Q) of an excited nuclear state at the probe nucleus and the electric field gradient (EFG) due to the surrounding charge distribution, thus being a unique tool to study the location of impurities, as well as their charge states and interaction with defects. EFG is a traceless diagonal tensor fully characterized by its $V_{zz}$ component and the axial asymmetry parameter $\eta = \qty(V_{xx}-V_{yy})/V_{zz}$, where the observable frequency is $\omega_0 \propto eQV_{zz}$ (see supplementary material).

\begin{figure}
	\centering
	\includegraphics[width=\linewidth]{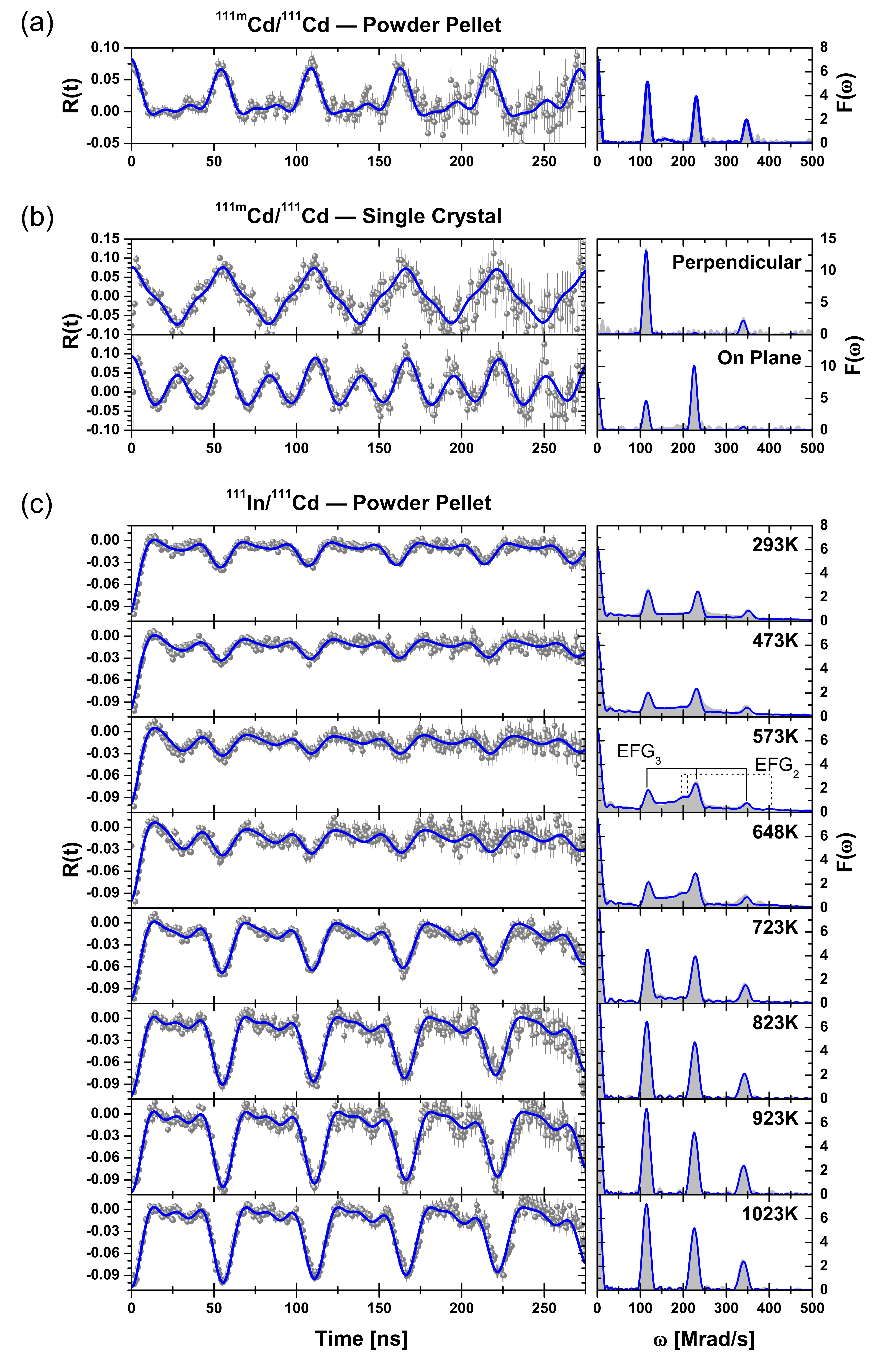}
	\caption{\label{fig:111In_111mCd_Ga2O3}PAC spectra (symbols) and their fits (lines) (as well as their Fourier transforms) after implantation of \ce{^{111m}Cd} \textbf{(a)} in the powder pellet and \textbf{(b)} in the single crystal, and \textbf{(c)} as a function of temperature after diffusion of \ce{^{111}In} in the powder pellet. The triplet of frequencies associated with the EFGs of the intermediate (2) and final (3) states are shown in the Fourier transform at $\SI{573}{\kelvin}$.}
\end{figure}

Two different radioactive impurity elements were used, \ce{^{111m}Cd (48min)} and \ce{^{111}In (2.8d)}, both decaying to stable \ce{^{111}Cd} by gamma-gamma ($\gamma$-$\gamma$) emission through the same 5/2+, 245 keV probe level of \ce{^{111}Cd} \cite{Nagl2013}. \ce{^{111m}Cd}, with no element transmutation, was used to directly study the location of the implanted dopants. It was implanted at room temperature in powder pellet and single crystal samples (see the supplementary material for details) at ISOLDE-CERN to a low fluence of \SI{e11}{atoms/cm^2}. After subsequent 10 minutes air annealing at \SI{1473}{\kelvin} and at \SI{1273}{\kelvin}, respectively, the measurements were carried out at room temperature using a standard 6-\ce{BaF2} detector PAC analog spectrometer \cite{Butz1989} (for the single crystal, considering two different surface normal orientations with respect to the detectors plane). For nuclear level spin 5/2, a triplet of frequencies characterizes each EFG at the PAC observable R(t) function and respective Fourier transform. The experiments show a dominant position for the Cd probes in the powder pellet with an EFG characterized by $\omega_0 = \SI{115.5}{Mrad/s}$, $\eta = \num{0.11}$ (Fig. \hyperref[fig:111In_111mCd_Ga2O3]{\ref*{fig:111In_111mCd_Ga2O3}(a)}), but $21\%$ of the probes were still in defective environment (probably due to grain boundaries) while the equivalent experiments done in the single crystal showed a similar EFG ($\omega_0 = \SI{112.9}{Mrad/s}$, $\eta = \num{0.09}$) with no remaining implantation defects (Fig. \hyperref[fig:111In_111mCd_Ga2O3]{\ref*{fig:111In_111mCd_Ga2O3}(b)}).

In the case of \ce{^{111}In}, the probe decays by electron capture of an inner atomic K- or L-shell electron, leaving \ce{^{111}Cd} in an ionized unstable state before the nuclear $\gamma$-$\gamma$ emission. The first inner hole created at the restructuring Cd atomic shells is very rapidly recovered (within about $10^{-14}$ s) by electrons of higher orbits with consequent emission of X-rays and/or Auger electrons leading to further ionization of the atom, losing on average 3 to 8 electrons \cite{Lupascu1996,Carlson1966}. While reaching stability, the charge distribution around the probe (hence the EFG) is changing, thus monitoring this EFG by PAC as a function of temperature, enables studying the formation (and annihilation) of ionized and excited electronic states of Cd. For the analysis of fluctuating EFGs, the theory of stochastic processes applied to dynamic transitions in PAC \cite{Winkler1973} is necessary, which assumes the environment around a probe might change with time in such a way that allowed transitions between different states occur in a "Markov chain"-like fashion. Therefore, the system is described by the EFGs that characterize each possible state and by the transition rates to go from one state to another. In this context, the experimental data was analyzed using a complex first principles fitting program integrating multiple EFG states and their mutual transition rates \cite{Barbosa2018}.

\begin{figure}
	\centering
	\includegraphics[width=\linewidth]{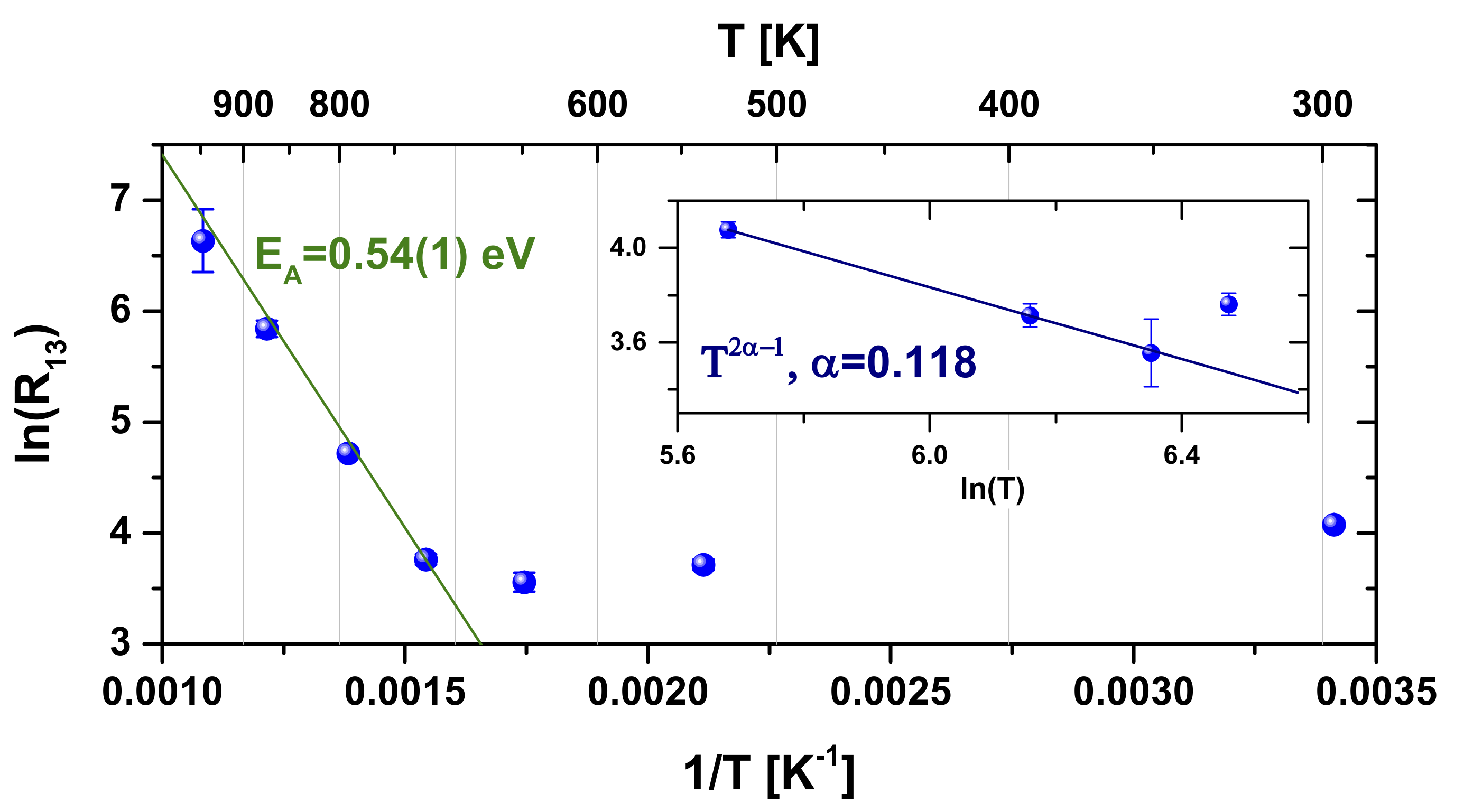}
	\caption{\label{fig:Ga2O3_Arrhenius}Arrhenius plot of the transition rate $R_{1 \rightarrow 3}$ (MHz) as a function of temperature in logarithmic scale and the corresponding fit between \SI{648}{\kelvin} and \SI{923}{\kelvin}. \textbf{(inset)} Logarithmic $R_{1 \rightarrow 3}$ plotted as a function of logarithmic temperatures between \SI{293}{\kelvin} and \SI{573}{\kelvin}. The linear curve corresponds to $T^{2\alpha -1}$ for $\alpha = 0.118$ (the slope of the curve is equal to $2\alpha -1$).}
\end{figure}

The powder pellet sample was wetted with a \ce{^{111}In} (aq) solution, dried and annealed in air during 48 hours at \SI{1373}{\kelvin} promoting \ce{^{111}In} diffusion. Then, PAC measurements were performed between \SI{293}{\kelvin} and \SI{1023}{\kelvin} (Fig. \hyperref[fig:111In_111mCd_Ga2O3]{\ref*{fig:111In_111mCd_Ga2O3}(c)}). Control room temperature (\SI{293}{\kelvin}) measurements were performed before and after the full series of experiments with no observable changes, ensuring that no irreversible annealing effects occurred. The spectra at \SI{923}{\kelvin} and at \SI{1023}{\kelvin} are indistinguishable with EFG signature characterized by $\omega_0 = \SI{113}{Mrad/s}$ and $\eta = \num{0.10}$, matching the EFG obtained from the \ce{^{111m}Cd} measurements. That same EFG is visible at all temperatures but the amplitude of the corresponding R(t) spectra clearly varies. Since the PAC measurements are reversible with temperature, and there are no structural $\beta$-\ce{Ga2O3} phase transitions in this range of temperatures \cite{Stepanov2016}, the observed effects must be due to the temperature dependence of the Cd electronic recovery after electron capture. Directly after the In decay, before sufficient electronic recovery is achieved, the charge distribution around a Cd atom is variable and uncertain. This leads to an initial state described by a broad EFG distribution with central EFG ($\omega_0 = \SI{111}{Mrad/s}$, $\eta = \num{1.0}$) and $\text{FWHM} > \SI{30}{Mrad/s}$. Then, a unidirectional transition occurs to a final stable state characterized by the EFG mentioned earlier. At high temperatures the electronic recovery is faster than the experimentally observable timescale, where only the final stable configuration is observed. Furthermore, all probes are found in a single location and the low damping $(\text{FWHM} = \SI{1.4}{Mrad/s})$ suggests that the concentration of point defects in the next-nearest lattice neighborhood are much below the $\sim$1 ppm probe's concentration. At \SI{573}{\kelvin} and \SI{648}{\kelvin} a meta-stable Cd state was also observed (intermediate state), characterized by $\omega_0 = \SI{117}{Mrad/s}$, $\eta = \num{0.92}$ and very low $\text{FWHM} < \SI{2}{Mrad/s}$. At these temperatures, this meta-stable electronic state of Cd has ${\sim}7\%$ probability of being formed before achieving the final state. Due to the large $\eta$, the first two observable triplet frequencies are very close, merging into a single broader peak around \SI{200}{Mrad/s} in the Fourier spectra. It was observed that the EFG of every state barely changes with temperature while the transition rates ($R_{i \rightarrow j}$) have strong variations, in particular for the transition $R_{1 \rightarrow 3}$ between the initial ($i=1$) and final ($j=3$) states above \SI{648}{\kelvin} (see Fig. \ref{fig:Ga2O3_Arrhenius}).

\begin{figure*}
	\centering
	\includegraphics[width=\linewidth]{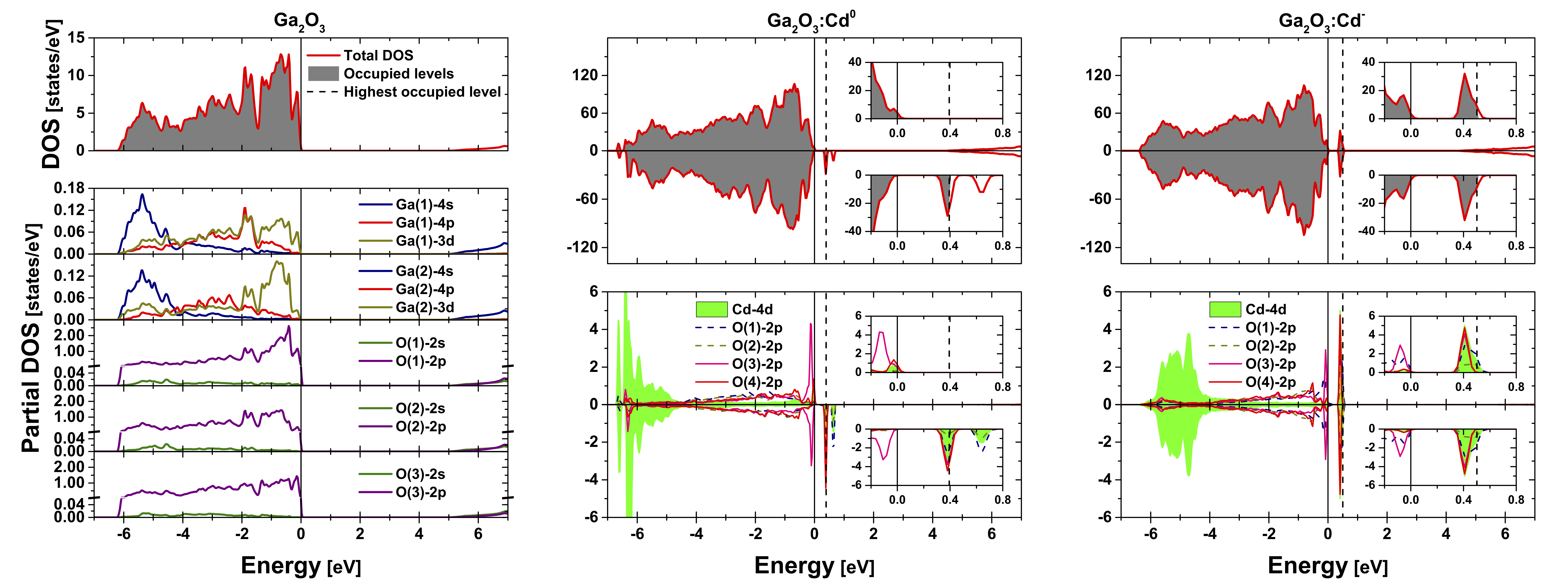}
	\caption{\label{fig:DOS}Total and atom-resolved partial density of states (DOS) of \ce{Ga2O3} and supercells of (\ce{Ga2O3}:Cd)\textsuperscript{0} and (\ce{Ga2O3}:Cd)\textsuperscript{-}. The top valence band of \ce{Ga2O3} is set at \SI{0}{\electronvolt} (full line) and the energy of the highest occupied level is represented by a dashed line. For pure \ce{Ga2O3}, both lines are coincident. The insets show a zoomed energy range focused on the impurity band. Partial DOS includes orbital contribution from all non-equivalent atoms of \ce{Ga2O3}, as well as the $4d$ contribution of \ce{Cd} and the $2p$ contribution of its neighboring \ce{O} atoms (due to symmetry break, \ce{Cd} has 4 non-equivalent \ce{O} neighbors instead of 3).}
\end{figure*}

The PAC results were compared to DFT calculations as implemented in the WIEN2k package \cite{WIEN2k} (see supplementary material for details).
A unit cell of $\beta$-\ce{Ga2O3} was considered using the lattice parameters reported by {\AA}hman \emph{et al.} \cite{Ahman1996}. It has a sixfold coordinated Ga site (octahedral), a fourfold coordinated Ga site (tetrahedral) and three non-equivalent O sites. After structural relaxation, a band gap of \SI{4.91}{\electronvolt} was obtained from the total density of states (DOS) (Fig. \ref{fig:DOS}) in good agreement with the experimental value of \SI{4.85}{\electronvolt} \cite{Stepanov2016}. Moreover, the atom-resolved partial density of states (PDOS) shows a very small contribution from the Ga-$3d$, Ga-$4s$ and Ga-$4p$ orbitals to the upper valence band that is completely dominated by the O-$2p$ orbitals, whereas the conduction band minimum is mainly described by the Ga-$4s$ orbitals (Fig. \ref{fig:DOS}).

From the occupancy of In atoms in similar \ce{InGaO3} and \ce{In2O3} structures \cite{Pasquevich1993}, it was expected that both In (diffused) and Cd (implanted) \cite{Barbosa2013} incorporate into the larger octahedral Ga site in \ce{Ga2O3}. Therefore, a \num{1x4x2} supercell of \ce{Ga2O3} was constructed having a Cd atom placed in an octahedral Ga site (1:64 dilution). Since Ga has 3 valence electrons whereas Cd has only 2, two different charge states were considered for the Cd probe: neutral (\ce{Cd^{0}}) and charged (\ce{Cd^{-}}), the latter by adding one extra electron to the lattice which is compensated by an additional homogeneous positive background to keep the entire cell in a neutral state \cite{Blochl1995, Darriba2012}. The EFGs at the Cd probe were calculated for each configuration (Table \ref{tab:Vzzs}) and a very good agreement is found between the EFGs for \ce{Cd^{-}} and the final state in the PAC experiments, as well as between the EFGs for \ce{Cd^{0}} and the intermediate state observed at \SI{573}{\kelvin} (Table S2 in the supplementary material contains the calculated EFGs for other Cd sites and charge states).

\begin{table}
	\centering
	\caption{\label{tab:Vzzs}$V_{zz}$ and $\eta$ for each simulated Cd charge state and PAC experimental states.}
	\begin{ruledtabular}
	\begin{tabular}{c c c}
		& $V_{zz}$ (\si{V/\angstrom^2}) & $\eta$ \\
		\hline
		\ce{Ga2O3}:\ce{Cd^{0}} & 70.2 & 0.98 \\
		\ce{Ga2O3}:\ce{Cd^{-}} & 63.8 & 0.04 \\
		\hline
		Exp. initial state & 64(2)\footnotemark[1] & 1.0(1) \\
		Exp. intermediate state & 67(2)\footnotemark[1] & 0.92(3) \\
		Exp. final state & 65(1)\footnotemark[1] & 0.10(1)
	\end{tabular}
	\end{ruledtabular}
	\footnotetext[1]{Using the electric quadrupole moment $Q = \SI{0.765+-0.015}{\barn}$ \cite{Haas2010}.}
\end{table}

\begin{table}
	\caption{\label{tab:Vzz_components}$p$, $d$ and $s$-$d$ valence contributions to the EFG tensor in the principal components axes (in units of \si{V/\angstrom^2}) for each Cd charge state (by definition, $\left| V_{zz} \right|\geq \left| V_{yy} \right| \geq \left| V_{xx} \right|$).}
	\begin{ruledtabular}
	\begin{tabular}{r c c c c c c}
		& \multicolumn{3}{c}{\ce{Ga2O3}:\ce{Cd^0}} & \multicolumn{3}{c}{\ce{Ga2O3}:\ce{Cd^{-}}} \\
		\cline{2-4} \cline{5-7}
		& $V_{xx}$ & $V_{yy}$ & $V_{zz}$ & $V_{xx}$ & $V_{yy}$ & $V_{zz}$ \\
		\hline
		p & -19.6 & -24.4 & 43.9 & -23.1 & -28.6 & 51.7 \\
		d & 18.3 & -46.0 & 27.7 & -7.9 & -5.0 & 12.9 \\
		s-d & 0.5 & 0.8 & -1.3 & 0.6 & 0.7 & -1.4
	\end{tabular}
	\end{ruledtabular}
\end{table}

The total DOS shows that replacing a Ga atom by a \ce{Cd^0} probe induces a partially filled impurity band in the band gap near the top valence band while the basic electronic structure of \ce{Ga2O3} remains unaltered (Fig. \ref{fig:DOS}). By integrating the unfilled region of the band, a value of 1.00 electron is obtained, in accordance with the expected behavior of a Cd dopant with single-acceptor character in this material. By looking at the partial DOS projected at the Cd and O atoms (Fig. \ref{fig:DOS}), it is possible to see that the impurity band is composed of Cd-$4d$ and O-$2p$ orbitals. A similar behavior is observed for \ce{Cd^{-}}, but in this case the impurity band gets completely filled. The top of the impurity band is $\sim$\SI{4}{\electronvolt} from the conduction band, which is very close to the \SI{4.3}{\electronvolt} reported for the band gap of amorphous Cd-Ga-O thin films \cite{Yanagi2015}. The analysis of the $p$, $d$ and $s$-$d$ valence contributions to the EFG tensor principal components ($V_{xx}, V_{yy}, V_{zz}$) for each Cd state (Table \ref{tab:Vzz_components}) reveals that the biggest variation occurs in the $d$ valence contribution, mainly in the $V_{xx}$ and $V_{yy}$ components. This is in perfect agreement with the DOS results, since the impurity band is composed of Cd-$4d$ states that get filled when going from the \ce{Cd^0} to the \ce{Cd^{-}} charge state. Consequently, relevant changes in $\eta$, which depends on $V_{xx}$ and $V_{yy}$, but little changes in $V_{zz}$ are expected for different charge states, in perfect agreement with both theoretical and experimental EFG values found for \ce{Cd^0} and \ce{Cd^{-}} (Table \ref{tab:Vzzs}).

Having identified the position and charge states of Cd, the temperature dependence of the transition rates between the initial (electronically unstable) and the final (stable) states in the \ce{^{111}In} PAC experiments provides insights about electron mobility in \ce{Ga2O3}, revealing two different regimes (Fig. \ref{fig:Ga2O3_Arrhenius}): \textbf{1)} Above \SI{648}{\kelvin}, the transition rates are described by an Arrhenius law with an activation energy of \SI{0.54+-0.01}{\electronvolt}. According to the present simulations, Cd induces an impurity band $\sim$\SI{0.4}{\electronvolt} above the top valence band of \ce{Ga2O3}, which is tempting to attribute to the experimental activation energy of \SI{0.54+-0.01}{\electronvolt}. However, \ce{^{111m}Cd} PAC measurements showed that, after annealing for removal of implantation defects, all Cd are found in the \ce{Cd^{-}} charge state at room temperature, suggesting that the Fermi level lies above the acceptor level, as expected for n-type material. Fleischer and Meixner \cite{Fleischer1993} reported a thermal activation energy of \SI{0.6+-0.1}{\electronvolt} for carrier mobility in \ce{Ga2O3} and associated it with electron delocalization and increasing electron mobility with increasing temperature. Nonetheless, Ma \emph{et al.} \cite{Ma2016} observed that charge carrier mobility and charge carrier density are interconnected, where ionized impurity scattering (which depends on donor concentration) decreases electron mobility. In addition, they identified polar optical phonon scattering as the dominant mechanism limiting electron mobility in $\beta$-\ce{Ga2O3} by observing a decrease in electron mobility with increasing temperature (in contradiction to the temperature-dependence observed by Fleischer and Meixner \cite{Fleischer1993}). Additionaly, deep level transient spectroscopy (DLTS) measurements performed by Irmscher \emph{et al.} \cite{Irmscher2011} showed the presence of several deep defect levels inside the band gap, including one \SI{0.55}{\electronvolt} below the conduction band and another \SI{0.74}{\electronvolt} below the conduction band (the second being dominant). Similar results were found by Zhang \emph{et al.} \cite{Zhang2016}, identifying defect levels \SI{0.62}{\electronvolt} and \SI{0.82}{\electronvolt} below the conduction band.
Therefore, at the light of what is known, the activation energy of \SI{0.54+-0.01}{\electronvolt} obtained in this work is most likely associated with electrons going from defect levels (observed using DLTS \cite{Irmscher2011,Zhang2016}) to the conduction band, recalling the mentioned interconnection between electron mobility and donor concentration \cite{Ma2016}.
\textbf{2)} Below \SI{573}{\kelvin}, a slight decrease of $R_{1 \rightarrow 3}$ with rising temperature is, unexpectedly, observed. This counterintuitive phenomenon cannot be explained by low temperature conductive processes such as variable-range hopping. However, similar behavior is predicted for the quantum tunneling rate in certain regimes of a biased double-well potential coupled to a dissipative field with ohmic dissipation (the coupling field has ohmic dissipation if its spectral density is proportional to the frequency: $J(\omega) \propto \omega$). For this reason, the hypothesis of direct electron tunneling from a defect to the Cd probe was explored assuming that the defect is in one of the potential wells and the Cd probe is in the other. The theory of such interaction has been developed in Refs. \cite{Chakravarty1984,Fisher1985,BirgeGolding_DefectTunneling} and applications can be found in the dynamics of single bistable defects coupled to an electron bath in disordered metals \cite{Chun1996} and in the tunneling of a Xe atom between the tip of a scanning-tunneling microscope (STM) or of an atomic-force microscope (AFM) and a Ni surface, coupled to phonons \cite{Louis1995}. According to this theory, when $k_BT \gg \epsilon$, where $\epsilon$ is the bias energy between the two wells, the tunneling rate $\gamma$ between the wells is approximately:

\begin{equation}
	\gamma \approx \frac{\Delta_r}{2} \qty(\frac{2\pi k_BT}{\hbar \Delta_r})^{2\alpha-1}\frac{\abs{\Gamma \qty(\alpha)}^2}{\Gamma(2\alpha)}
	\label{eq:tunneling_rate_highT}
\end{equation}

\begin{eqnarray}
	\label{eq:Ln_tunneling_rate_highT}
	\ln(\gamma) & \approx & \ln(\theta) +\qty(2\alpha -1)\ln(T), \\ 
		\theta & = & \frac{\Delta_r}{2} \qty(\frac{2\pi k_B}{\hbar \Delta_r})^{2\alpha-1}\frac{\abs{\Gamma \qty(\alpha)}^2}{\Gamma(2\alpha)} \nonumber
\end{eqnarray}

with $\Delta_r=\Delta_0 \qty(\frac{\Delta_0}{\omega_p})^\frac{\alpha}{1-\alpha}$, where $\omega_p$ is the oscillation frequency in either well, $\Delta_0$ is the bare tunneling matrix element, $\Gamma(x)$ is the Gamma function and $\alpha$ is a dimensionless dissipation coefficient which is dependent on the material. In the regimes where $0 < \alpha < 1/2$, the tunneling rate increases with decreasing temperature. Fitting the logarithmic equation to the transition rates obtained between \SI{293}{\kelvin} and \SI{573}{\kelvin} yields $\alpha = \num{0.118+-0.004}$ and $\Delta_r = \SI{1.3(2)e10}{\per\second}$ (see inset of Figure \ref{fig:Ga2O3_Arrhenius}), hence we are in fact in the regime of $0 < \alpha < 1/2$ and our transition rates behave as predicted by equation \ref{eq:tunneling_rate_highT}. The theory is only valid for a specific set of values for each variable, so we can test them and set boundaries in our case; firstly, $k_BT > \epsilon$, thus $\epsilon < \SI{0.0252}{\electronvolt}$. Then, the theory requires that $\hbar \Delta_r \ll \alpha k_BT$, which is true even at room temperature $\qty(\SI{8.53e-6}{\electronvolt} \ll \SI{2.98e-3}{\electronvolt})$. The last requirement is that $\Delta_0 \ll \omega_p$. Considering as reference the frequency associated with the polar optical phonon energy reported in Ref. \cite{Ma2016} $\qty(\omega_p=\SI{6.7(6)E13}{\per\second})$ and using the definition of $\Delta_r$, we have $\Delta_0 = \SI{3.6(6)E10}{\per\second}$, therefore $\Delta_0 \ll \omega_p$ is verified. A hypothesis for the observations at lower temperatures can then be formulated, stating that electrons are tunneling from some defects directly to the Cd probes and that such a system can be described by a double-well potential coupled to a dissipative field with ohmic dissipation. In this case, we believe that phonons are the dissipative field, as observed in the example above from Louis et al. \cite{Louis1995}. Although Leggett \emph{et al.} \cite{Leggett1987,Leggett1987errata} suggest a super-ohmic dissipation ($J(\omega) \propto \omega^s$, with $s > 1$) in the case of defect (or electron) tunneling in a solid with coupling to a (three-dimensional) acoustic phonon bath, in our understanding the authors refer to diffusion by quantum tunneling, whereas in our case there is solely direct electron tunneling between a specific defect and a Cd probe, thus the possibility of ohmic dissipation remains acceptable. Therefore, we propose that the lower temperature regime of the observed electronic recovery is explained by direct electron tunneling from defects to the Cd probes.

Concerning the issue of strong hole localization in \ce{Ga2O3} referred previously \cite{Varley2012}, since no holes are present around the Cd probes, no polarons are formed (they would be readily observable in the PAC measurements). Besides, the Cd-O bonds showed a more covalent character, in contrast to the ionic character observed for the Ga-O bonds (Fig. S3 in the supplementary material), so the probability of polaron formation in an adjacent O atom is reduced.

Lastly, in the absence of experimental values, the ionization energy of the Cd acceptor has been estimated in previous DFT studies \cite{Lyons2018,Peelaers2019} suggesting a thermodynamic transition level for \ce{Cd^{0}}/\ce{Cd^{-}} in \ce{Ga2O3} with a deep level character ($\sim$\SI{1}{\electronvolt}). If such level is confirmed experimentally, Cd would be unsuitable as acceptor for efficient p-type doping. However, the \textit{ad hoc} numerical procedure there employed to take into account the Coulomb potentials of charged defects has been questioned by Wu \emph{et al.} \cite{Wu2017}, eventually leading to overestimated values. Moreover, in this work, that same method employed with a different functional suggests a transition of $\sim$\SI{0.4}{\electronvolt}.

In summary, we have shown that Cd occupies a single position in $\beta$-\ce{Ga2O3}, the substitutional octahedral Ga site, with a clear acceptor character. Two Cd charge states were observed, the neutral unstable \ce{Cd^0} with a short mean life time ($\frac{1}{R_{23}} \approx \SI{2}{\micro\second}$ at \SI{573}{\kelvin}, see the supplementary material) and the stable ionized acceptor \ce{Cd^{-}}. Two regimes where observed with distinct transport properties: above \SI{648}{\kelvin} the electron mobility and/or density increases with increasing temperature with an activation energy of \SI{0.54+-0.01}{\electronvolt}. For lower temperatures, surprisingly the local conductivity seems to decrease with increasing temperature which is attributed to an electron tunnelling mechanism. As a final note, it was shown that the use of very accessible \ce{^{111}In} activated solutions (currently used in hospitals as tracers) might be a feasible option to properly dope a $\beta$-\ce{Ga2O3} sample with Cd by transmutation.

\section*{Supplementary Material}
See Supplementary Material for further details on the PAC experimental technique, sample preparation, fitting parameters and DFT simulations, which
includes Refs. \cite{Schatz1996,Villora2004,Litimein2009,Tran2009,Kim2010,Mohamed2010,Setyawan2010,Koller2011,Peelaers2015}.

\begin{acknowledgments}
This work was performed within the ISOLDE proposal IS481 and supported by FCT-Portugal, projects CERN-FP-123585-2011, CERN-FIS-PAR-0005-2017 and PTDC/CTM-CTM/28011/2017, LISBOA-01-0145-FEDER-028011, and by the European Commission through FP7- ENSAR (contract 262010) and Horizon 2020 program ENSAR2 (contract 654002). M. B. B. acknowledges a scholarship from FCT, SFRH/BD/97591/2013, J. S. a grant from the Federal Ministry of Education and Research (BMBF), 05K16PGA. The authors further acknowledge E. G. V\'{i}llora and K. Shimamura (NIMS, Japan) for supplying the single crystal samples and the ISOLDE-CERN collaboration for supportive access to beam time.
\end{acknowledgments}

\nocite{Schatz1996,Villora2004,Litimein2009,Tran2009,Kim2010,Mohamed2010,Setyawan2010,Koller2011,Peelaers2015}


\providecommand{\noopsort}[1]{}\providecommand{\singleletter}[1]{#1}%

\end{document}


\title{Supplementary Material for\\ ``Cd acceptors in \ce{Ga2O3}, an atomistic view''}

\author{M. B. Barbosa, J. G. Correia, K. Lorenz, A. S. Fenta, J. Schell, R. Teixeira, E. Nogales, B. M\'{e}ndez, A. Stroppa, J. P. Ara\'{u}jo}

\maketitle

\renewcommand{\thefigure}{S\arabic{figure}}
\setcounter{figure}{0}

\renewcommand{\thetable}{S\arabic{table}}
\setcounter{table}{0}

\section{Perturbed Angular Correlations}
\subsection{Theoretical Background}

In a Perturbed Angular Correlation experiment, a radioactive probe which decays in a double cascade (emitting two photons, $\gamma_1$ and $\gamma_2$) is introduced in a sample by implantation, diffusion or neutron activation. The hyperfine interaction of the electric field gradient (EFG) at the probe's site with the electric quadrupole moment of the intermediate level of the cascade causes a time-dependent perturbation in the angular dependence of the emission probability of $\gamma_2$ with respect to $\gamma_1$. Since the EFG is a traceless matrix and diagonal in its principal axis, it can be completely described by only two parameters: the $V_{zz}$ component and the axial asymmetry parameter $\eta = \qty(V_{xx}-V_{yy})/V_{zz}$, considering that $\qty|V_{xx}| \leq |V_{yy}| \leq |V_{zz}|$ \cite{Schatz1996,Wichert1999}.

The time-dependent oscillations in the anisotropic emission of $\gamma_2$ then define the observable frequency $\omega_0$ which is proportional to the quadrupole interaction frequency $\omega_Q$

\begin{equation}
	\omega_0 = k\omega_Q, \quad \omega_Q = \frac{eQV_{zz}}{4I(2I-1)\hbar}
\end{equation}

where $I$ and $Q$ are the spin and the electric quadrupole moment of the intermediate level of the cascade, respectively, and $k=3$ (or 6) for integer (or half-integer) spin \cite{Schatz1996,Wichert1999}.

The coincidence spectra $N(\theta,t)$ can be recorded, where $\theta$ is the angle between detectors and $t$ is the time delay between the detection of $\gamma_1$ and $\gamma_2$, allowing for the experimental perturbation function

\begin{equation}
	R(t)=2\frac{N(\ang{180},t)-N(\ang{90},t)}{N(\ang{180},t)+2N(\ang{90},t)}\approx\sum A_{kk}G_{kk}
\end{equation}

to be calculated, where $A_{kk}$ are the angular correlation coefficients of the nuclear decay cascade and $G_{kk}$ is the perturbation factor. For a polycrystalline sample, $G_{kk}=S_{k0}+\sum_nS_{kn}\cos{\omega_nt}$, so it is described by a sum of oscillatory terms with frequencies $\omega_n$ which correspond to transitions between the hyperfine levels created due to the splitting of the nuclear energy levels by the hyperfine interaction. The splitting of an intermediate level with spin $I=5/2$ results in three sub-levels, so transitions between them will yield a triplet of frequencies $\omega_1$, $\omega_2$ and $\omega_3=\omega_1+\omega_2$, with $\omega_n=C_n(\eta)\omega_0$. Therefore, for each EFG present in the system, three peaks will be observed in the Fourier transform of the perturbation function.

In the case of fluctuating EFGs, such as during the recovery processes of the electronic environment after the loss of electrons from the lower atomic orbits due to electron capture decay, a dynamic description of PAC spectra is needed. Here, the theory based on stochastic processes applied to PAC developed by Winkler and Gerdau \cite{Winkler1973} was considered. In their formalism, the time-evolution operator changes from the usual $\hat{\Omega}(t)=\exp \qty(-\frac{i}{\hbar}\hat{H}t)$ to $\hat{\Omega}(t)=\exp \qty[\qty(-\frac{i}{\hbar}H_{st}^{\times}+\hat{R})t]$, where $\qty(-\frac{i}{\hbar}H_{st}^{\times}+\hat{R})$ is called the Blume matrix and $H_{st}^{\times}$ and $\hat{R}$ are Liouville operators, the former being constructed from the Hamiltonians that describe the different possible states (each one described by a static EFG) and the latter containing the transition rates between the different possible states (where the inverse of the sum of all transition rates from one state to the others corresponds to the mean life of that state). In this case, the observable perturbation factor is given by

\begin{equation}
	G_{kk}(t)=\sum_{q=1}^{\qty(2I+1)^2N} a_{kq} \cos (\omega_q t) \exp (-\lambda_q t)
\end{equation}

where $N$ is the number of possible states, the amplitudes $a_{kq}$ depend on the eigenvectors of the Blume matrix and $-\lambda_q+i \omega_q$ are the eigenvalues. The real components of the eigenvalues are always negative and are only non-zero when the transition rates are also non-zero, thus the damping that they induce in the oscillations of the perturbation function are a signature of the presence of dynamic processes.

In this work, $N=3$ different states were needed to describe the \ce{^{111}In} PAC experimental data, so the system is described by the $\omega_0$, $\eta$ and initial percentage at $t=0$ of each individual state, plus the 6 transition rates between all pairs of states ($1 \rightarrow 2$, $1 \rightarrow 3$, $2 \rightarrow 1$, etc.), all of them acting as fitting parameters.
Moreover, to account for the fact that probes in equivalent sites might have slight deviations in their EFGs (e.g., due to possible remaining diluted defects not preferentially attached to them), a Lorentzian distribution characterized by their central frequency $\omega_0$ and full width at half maximum (FWHM) is integrated for each EFG individually in the fitting function.

The fitting is done by minimizing a chi-square function and the error of each fitting parameter is assumed to be the amount that they have to change in order for the chi-square function to vary one standard deviation.

\subsection{Experimental Details}

 99.999\% purity \ce{Ga2O3} powder was pressed into pellets of about \SI{7}{\milli\meter} diameter and \SI{2}{\milli\meter} thickness and subsequently annealed at \SI{11773}{\kelvin} for 8 h. Single crystals were grown by the floating zone technique using 4N purity powder, then cut and polished in the (1 0 0) plane, as described elsewhere \cite{Villora2004}. \ce{^{111}In} probes were introduced by wetting the powder pellet in a \ce{^{111}In} activated solution and annealed for 48 hours at \SI{1373}{\kelvin} in air. The PAC measurements were performed as a function of the measurement temperature (between \SI{293}{\kelvin} and \SI{1023}{\kelvin}) in a 4-\ce{BaF2} detector spectrometer \cite{Butz1989} at CFNUL, in Lisbon.
 \ce{^{111m}Cd} probes were implanted in a powder pellet sample and in a single crystalline sample at ISOLDE/CERN to low fluences of \SI{e11}{atoms/cm^2} at room temperature. Then, the samples were annealed for 10 minutes at \SI{1473}{\kelvin} and at \SI{1273}{\kelvin} in air, respectively, in order to remove implantation defects. The PAC measurements were carried out at room temperature on a 6-\ce{BaF2} detector spectrometer \cite{Butz1989}. For the single crystal, two orientations were considered in a 4-detector's plane: surface normal perpendicular to the detector's plane and in-plane at \ang{45} from the detectors.
 
\begin{figure}[!t]
	\centering
	\includegraphics[width=\linewidth]{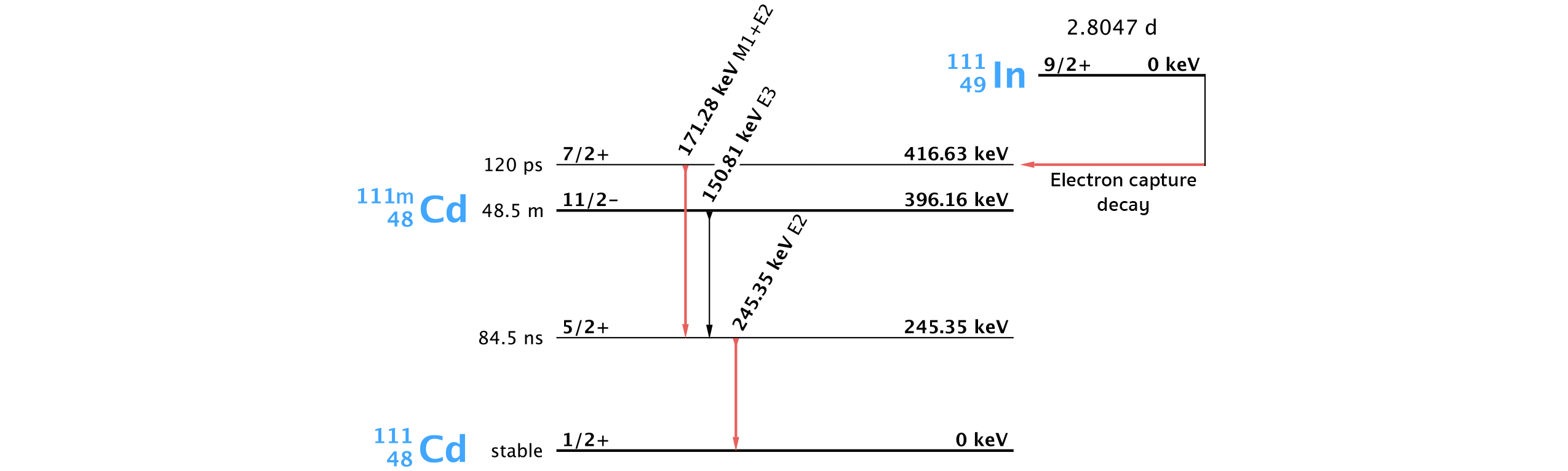}
	\caption{Decay scheme for \ce{^{111}In} and \ce{^{111m}Cd}}
	\label{fig:111In_111Cd_Cascades}
\end{figure}

 The decay scheme of both \ce{^{111}In} and \ce{^{111m}Cd} can be seen in Figure \ref{fig:111In_111Cd_Cascades} and the fitting parameters from \ce{^{111}In} PAC as a function of temperature are summarized in Table \ref{tab:111In_Ga2O3_poly}.
 
\begin{table}[!htb]
	\centering
	\caption{\label{tab:111In_Ga2O3_poly}\ce{^{111}In}:\ce{Ga2O3} PAC fitting parameters. Three different states described by a single EFG each were considered. The transition rates between them are expressed in MHz, the quadrupole frequency $\omega_0$ and the full width at half maximum for the Lorentzian, static-like, distribution (FWHM) are expressed in Mrad/s and the asymmetry parameter $\eta$ is dimensionless.}
	\begin{ruledtabular}
	\begin{tabular}{c c c c c c c c c c c c c}
 		\multirow{2}{*}{T (\si{\kelvin})} & \multicolumn{3}{c}{State 1 (initial)} & \multicolumn{3}{c}{State 2 (intermediate)} & \multicolumn{3}{c}{State 3 (final)} & \multicolumn{3}{c}{Transition Rates} \\
 		\cline{2-4} \cline{5-7} \cline{8-10} \cline{11-13}
  		& $\omega_0$ & $\eta$ & FWHM & $\omega_0$ & $\eta$ & FWHM & $\omega_0$ & $\eta$ & FWHM & $1\rightarrow2$ & $1\rightarrow3$ & $2\rightarrow3$ \\
		\hline
		293 & \multirow{7}{*}{111(1)} & \multirow{7}{*}{1.0(1)} & 79(2) & - & - & - & 117.8(1) & 0.000(1) & \multirow{7}{*}{1.4(3)} & - & 59(2) & - \\
		473 & & & 66(2) & - & - & - & 116.4(2) & 0.000(2) & & - & 41(2) & - \\
		573 & & & 47(8) & \multirow{2}{*}{117(2)} & \multirow{2}{*}{0.92(3)} & \multirow{2}{*}{0(2)} & 116.1(1) & 0.000(1) & & 3(1) & 35(5) & 0.5(1) \\
		648 & & & 34(2) & & & & 115.7(2) & 0.000(5) & & 3(2) & 43(2) & 8(4) \\
		723 & & & 20(2) & - & - & - & 114.6(1) & 0.000(3) & & - & 112(3) & - \\
		823 & & & 25(6) & - & - & - & 113.7(1) & 0.09(1) & & - & 344(26) & - \\
		923 & & & 0(2) & - & - & - & 113.1(1) & 0.10(1) & & - & 761(216) & - \\
	\end{tabular}
	\end{ruledtabular}
\end{table}

\section{Density Functional Theory simulations}

\subsection{Simulation Details}

The simulations were performed via the full-potential (linearized) augmented plane wave plus local orbitals [FP-(L)APW+lo] method as implemented in the WIEN2k code \cite{WIEN2k}. For the structural optimization and calculation of the electric field gradients, the generalized gradient approximation in the Perdew, Burke and Ernzerhof parameterization (GGA-PBE) \cite{WIEN2k} was considered as exchange-correlation functional. For the calculation of the density of states, band structure and band gap, the optimized structures were used and the modified Becke-Johnson exchange potential (mBJ) was applied, since it has been proven to better estimate band gaps in semiconducting materials than simply using GGA \cite{Tran2009,Koller2011}. It also has levels of agreement with experimental results comparable to hybrid functionals or Green function (GW) methods (which are computationally heavier and more time consuming) whilst being barely more expensive than GGA calculations \cite{Tran2009,Koller2011}. Although hybrid functionals have been previously proven very successful to simulate \ce{Ga2O3} systems \cite{Varley2010,Varley2012}, they take three to four orders of magnitude more time in a simulation than GGA, thus being impractical for the several big supercells required in the present study.

The structural parameters of \ce{Ga2O3} in the $\beta$-phase ($\beta$-\ce{Ga2O3}) as found in the work of {\AA}hman \emph{et al.} \cite{Ahman1996} were considered, i.e. $a = \SI{12.214+-0.003}{\angstrom}$, $b = \SI{3.0371+-0.0009}{\angstrom}$, $c = \SI{5.7981+-0.0009}{\angstrom}$, $\alpha = \gamma = \ang{90}$ and $\beta = \ang{103.83+-0.02}$, with the internal atomic positions being optimized by minimizing the atomic forces to a maximum limit of \SI{2}{mRy/bohr} in a self-consistent way. Optimization of the lattice parameters using the very precise HSE06 hybrid functional was previously reported elsewhere \cite{Peelaers2015} and the calculated lattice parameters are very close to the experimental ones (less than 0.5\% variation), therefore no lattice optimization was performed in this work and the experimental values were used for the simulations.

To simulate an isolated Cd impurity, a \num{1x4x2} supercell of \ce{Ga2O3} with dimensions $a' = a = \SI{12.214}{\angstrom}$, $b' = 4b = \SI{12.1484}{\angstrom}$, $c' = 2c = \SI{11.5962}{\angstrom}$ and $\beta = \ang{103.83}$ was constructed. Its size was determined by increasing it until the variation of the EFG at the Cd site was in the same order of magnitude of the PAC experimental error.

A cut-off value for the plane wave expansion of Rmt*Kmax = 6.0 was considered, where Rmt is the muffin-tin sphere radius and Kmax is the largest K-vector of the plane wave expansion of the wave function. 90 and 20 k-points in the irreducible Brillouin zone were used for the \ce{Ga2O3} simple cell and for the \num{1x4x2} supercell with the Cd impurity, respectively.

Different charge states for the Cd probes were considered, where additional charges were compensated by adding a homogeneous background of opposite charge to keep the entire cell in a neutral state \cite{Blochl1995,WIEN2k,Darriba2012}. For example, if an electron is added to the cell, the extra negative charge can be localized but a uniform positive charge will maintain the neutrality of the cell whilst not resulting in any extra interactions.

For the estimation of the thermodynamic transition level for Cd0/Cd-1, the procedure employed in Refs. \onlinecite{Lyons2018} and \onlinecite{Peelaers2019} was used, but the energy alignment in relation to bulk \ce{Ga2O3} was performed using the core energy levels from atoms far from the Cd probes instead of using the electrostatic potential.


\subsection{Band Structure}

\begin{figure}[htb]
	\centering
	\includegraphics[width=\linewidth]{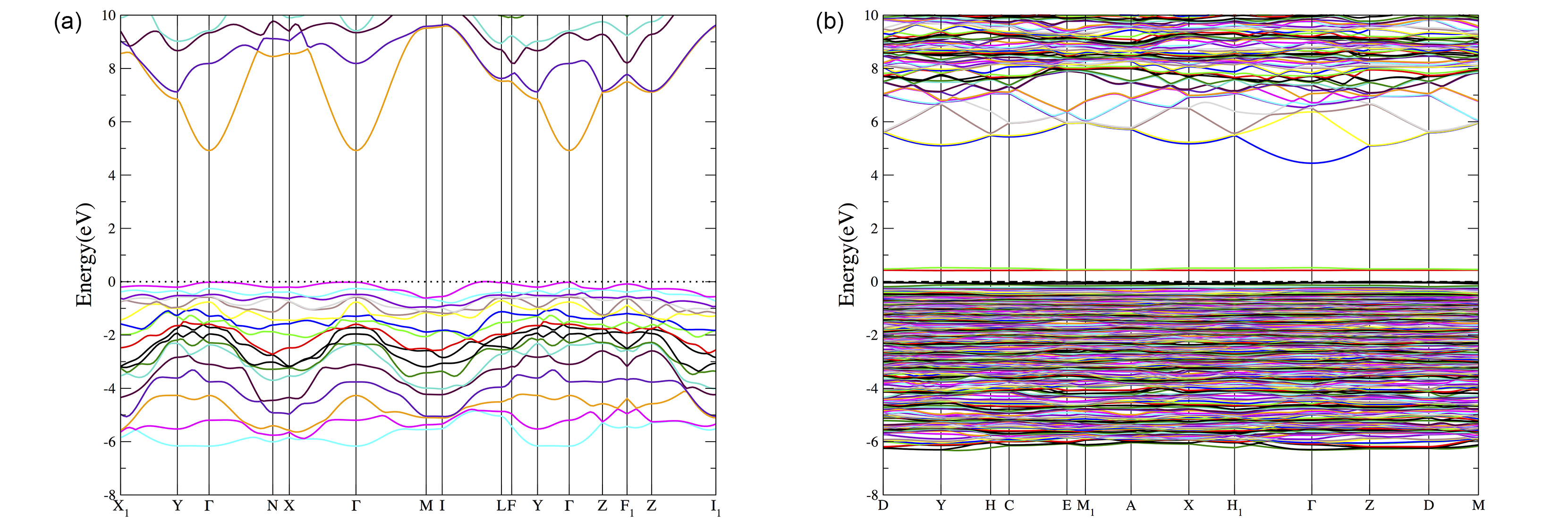}
	\caption[Band structure of \ce{Ga2O3} and supercell with \ce{Cd^{-}} in an octahedral \ce{Ga} site]{Band structure of \textbf{(a)} pure \ce{Ga2O3} and \textbf{(b)} \ce{Cd^{-}} in an octahedral \ce{Ga} site of a \num{1x4x2} supercell. The top valence band of pure \ce{Ga2O3} is set at \SI{0}{\electronvolt}. The k-point labels are named as in Ref.\cite{Peelaers2015} for pure \ce{Ga2O3} and as in Ref.\cite{Setyawan2010} for the supercell containing \ce{Cd}.}
	\label{fig:BandStructureGa2O3}
\end{figure}

The band structure of \ce{Ga2O3} (Fig. \ref{fig:BandStructureGa2O3}) shows an indirect band gap between the valence band maximum (VBM) located on the I-L line and the free-electron-like conduction band minimum (CBM) at the Gamma point. However, the valence band at the Gamma point is only \SI{0.03}{\electronvolt} below that of the VBM, so there is an indirect band gap of \SI{4.91}{\electronvolt} and a direct band gap of \SI{4.94}{\electronvolt}, in good agreement with optical absorption measurements \cite{Tippins1965} and with previous calculations \cite{Ratnaparkhe2017,Peelaers2015}. By fitting the energy dispersion of the CBM at the $\Gamma$ point to a parabolic function, an electron effective mass ($m^{*}_e$) of 0.35 $m_e$ was obtained. This is close to the experimental value of 0.28 $m_e$ \cite{Mohamed2010} but slightly higher, which is expected since the used mBJ exchange potential generally overestimates the effective masses \cite{Kim2010}. On the other hand, the top valence band is almost flat, indicating a rather large effective mass ($m^{*}_h$) for holes. This suggests that the electronic conductivity in \ce{Ga2O3} strongly depends on the mobility of the electrons that are thermally excited to the conduction band and less on the movement of holes created at the same time. These results are consistent with previous reports \cite{Ratnaparkhe2017,Varley2010,Varley2012,Peelaers2015,Litimein2009}.

In the band structure of the \num{1x4x2} supercell with \ce{Cd^{-}} in an octahedral \ce{Ga} site (Fig. \ref{fig:BandStructureGa2O3}(b)), it is possible to see the induced impurity band (which is $\sim$\SI{0.4}{\electronvolt} above the top of the valence band) and that the top of the valence band remains very flat as in pure \ce{Ga2O3}, thus the effective mass for holes remains very large.

\subsection{Electron Density}

\begin{figure}[htb]
	\centering
	\includegraphics[width=0.9\linewidth]{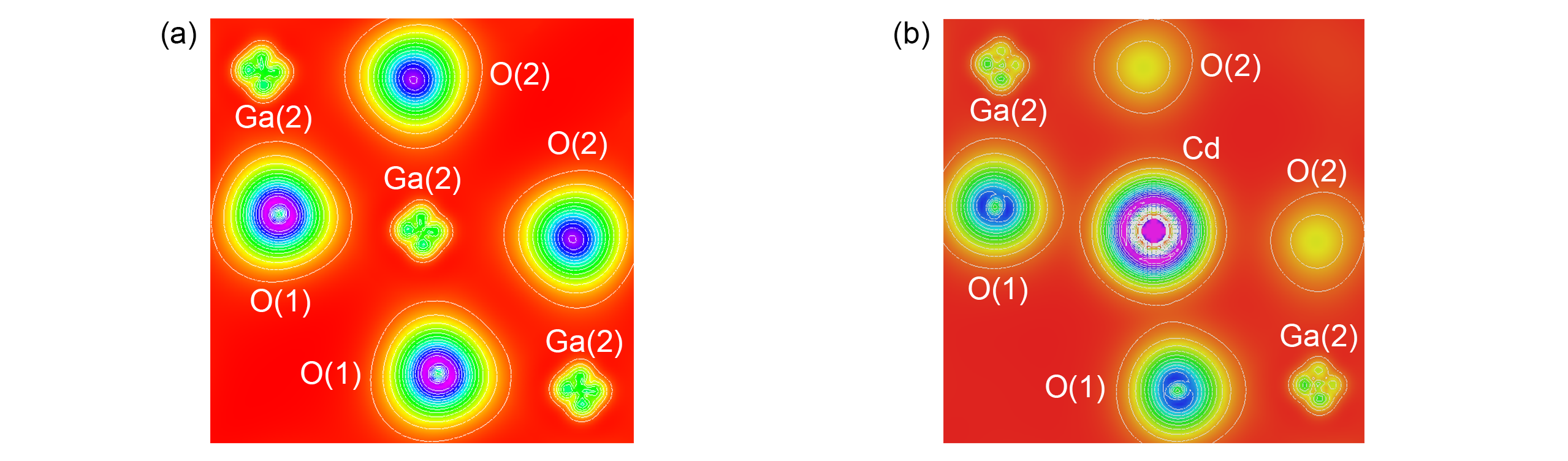}
	\caption[Electron density in pure \ce{Ga2O3} and in \ce{Cd^{-}} in an octahedral \ce{Ga}]{Electron density considering the top of the valence band in a plane going through the centre of \textbf{(a)} Ga(2) and O(1) in pure \ce{Ga2O3} simple cell and of \textbf{(b)} O(1) and \ce{Cd^{-}} in an octahedral \ce{Ga} site within a \num{1x4x2} supercell.}
	\label{fig:ElectronDensityGa2O3}
\end{figure}

The electron density of \ce{Ga2O3} (Fig. \ref{fig:ElectronDensityGa2O3}(a)) shows that the electron density is highest around the O atoms (similar picture is observed in any cut-plane direction) thus hinting to an ionic-like character for the Ga-O chemical bonds.

In the supercell case containing Cd (Fig. \ref{fig:ElectronDensityGa2O3}(b)), it is possible to see that the charges are distributed between the Cd atom and its O atomic neighbors, indicating that the Cd-O bonds have a more covalent character in contrast to the ionic character exhibited by the Ga-O bonds for the Ga atoms in the same position.

\subsection{Electric Field Gradient}

\begin{figure}[!htb]
	\centering
	\includegraphics[width=\linewidth]{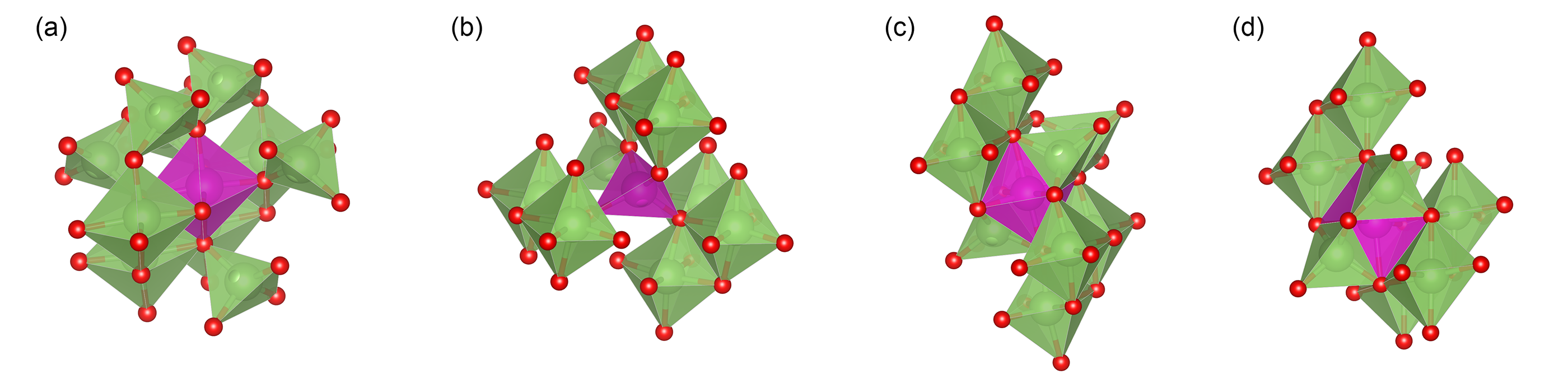}
	\caption[Simulated Cd sites in supercells of \ce{Ga2O3}]{Simulated Cd sites (in magenta) in supercells of $\beta$-\ce{Ga2O3}: \textbf{(a)} octahedral Ga site, \textbf{(b)} tetrahedral Ga site, \textbf{(c)} interstitial (I) and \textbf{(d)} interstitial (II). The figures do not show the entire calculated supercells.}
	\label{fig:Ga2O3_Cd_all_sites}
\end{figure}

Besides the EFGs calculated for each Cd probe's charge state at the octahedral Ga site reported in the letter as matching the PAC experimental results, the EFGs for Cd probes at other sites and charge states were calculated as well. Each considered site is represented in Figure \ref{fig:Ga2O3_Cd_all_sites} and the resulting EFGs are gathered in Table \ref{tab:All_Vzzs}.

\begin{table}
	\centering
	\caption{\label{tab:All_Vzzs}Calculated $V_{zz}$ and $\eta$ for the Ga atoms in a simple cell of $\beta$-\ce{Ga2O3} and for each site (see Fig. \ref{fig:Ga2O3_Cd_all_sites}) and charge state of the Cd probe in a \num{1x4x2} supercell. The calculated distance to the oxygen nearest neighbors (ONN) is also shown.}
	%
	\begin{ruledtabular}
	\begin{tabular}{l c c c}
		& $V_{zz}$ (\si{V/\angstrom^2}) & $\eta$ & ONN (\si{\angstrom}) \\
		\hline
		Ga \--- octahedral & 23.8 & 0.34 & 1.92 \--- 2.04 \\
		Ga \--- tetrahedral & -38.4 & 0.72 & 1.83 \--- 1.85 \\
		\hline
		\textbf{\ce{Cd^{0}}} \textbf{ \--- octahedral} & \textbf{70.2} & \textbf{0.98} & \textbf{2.11 \--- 2.25} \\
		\textbf{\ce{Cd^{-}}} \textbf{ \--- octahedral} & \textbf{63.8} & \textbf{0.04} & \textbf{2.13 \--- 2.32} \\
		\ce{Cd^{0}} \--- tetrahedral & -44.0 & 0.75 & 2.04 \--- 2.08 \\
		\ce{Cd^{-}} \--- tetrahedral & -52.6 & 0.59 & 2.06 \--- 2.10 \\
		\ce{Cd^{0}} \--- interstitial (I) & 38.7 & 0.45 & 2.17 \--- 2.20 \\
		\ce{Cd^{2+}} \--- interstitial (I) & 45.2 & 0.49 & 2.16 \--- 2.18 \\
		\ce{Cd^{0}} \--- interstitial (II) & -171.4 & 0.40 & 2.08 \--- 2.27 \\
		\ce{Cd^{2+}} \--- interstitial (II) & -195.6 & 0.95 & 2.00 \--- 2.59 \\
	\end{tabular}
	\end{ruledtabular}
\end{table}

\providecommand{\noopsort}[1]{}\providecommand{\singleletter}[1]{#1}%